\documentclass{aastex631}

\usepackage{comment}
\usepackage{listings}
\usepackage{graphicx} 
\usepackage{amssymb,amsmath,url}

\newcommand{\chisq}{$\chi^2$}

\newcommand{\V}{\mathbf V}
\newcommand{\A}{\mathbf A}
\newcommand{\I}{\mathbf I}

\newcommand{\FPW}{{\it FPW}}
\newcommand{\SFPW}{{$S_{FPW}$}}

\begin{document}
 
\title{A Fast Periodicity Detection Algorithm\\
Sensitive to Arbitrary Waveforms}

\author{Douglas P. Finkbeiner}
\affiliation{Department of Physics, Harvard University, Cambridge, MA 02138, USA}
\author{Thomas A. Prince}
\affiliation{Division of Physics, Mathematics and Astronomy, California Institute of Technology, Pasadena, CA 91125, USA}
\author{Samuel E. Whitebook }
\affiliation{Division of Physics, Mathematics and Astronomy, California Institute of Technology, Pasadena, CA 91125, USA}

\correspondingauthor{Thomas A. Prince}
\email{prince@caltech.edu}

\begin{abstract}
    A reexamination of period finding algorithms is prompted by new
    large area astronomical sky surveys  that can identify billions of individual sources having a thousand or more observations per source.
    This large increase in data necessitates
    fast and efficient period detection algorithms.
    In this paper, we provide an initial description of an algorithm that is being used for detection of periodic behavior
    in a sample of 1.5 billion objects using light curves generated from Zwicky Transient Facility (ZTF) data \citep{2019BellmPASP,Masci_ZTF_Survey}. 
    We call this algorithm ``Fast Periodicity Weighting'' (FPW), derived using a Gaussian Process (GP) formalism.
    A major advantage of the FPW algorithm for ZTF analysis is that it is agnostic to the details
    of the phase-folded waveform.  Periodic sources in ZTF show a wide variety of waveforms,
    some quite complex, including eclipsing objects, sinusoidally varying objects also exhibiting eclipses, objects with cyclotron emission at various phases, and accreting
    objects with complex waveforms. 
    We describe the FPW algorithm and its application to ZTF, and provide efficient code for both CPU and GPU.
\end{abstract}


\section{Introduction}
Period finding plays an important role in modern astronomy. The considerable increase of optical photometric survey data over the last two decades (see, e.g., \citet{Rau_PTF_Survey,2009LawPASP, 2019BellmPASP, Drake_CRTS,2016PanSTARRSsarXiv,2015RickerJATIS,2016GaiaA&A,2016DECamMNRAS,2018TonryPASP,2018JayasingheMNRAS}), combined with new and upcoming synoptic sky surveys (see, e.g., \citet{Ivezic_LSST,2016MeerlichtBlackgemSPIE,2022LS4arXiv,2023WSFTRAA}), places growing emphasis on periodicity detection algorithms that are fast and model agnostic, as well as those with the potential to combine data from different optical filters \citep{vanderplas2015multiband}. The most common periodograms seek to use least-squares analysis to fit a set of periodic basis functions to data. 
These include algorithms which fit sinusoidal basis functions (e.g. Lomb-Scargle \citep{Lomb_Periodogram, 
VanderPlas_UnderstandingLS} and the sinusoid variant of the Analysis of Variance (AOV) algorithm \citep{1996SchwarzApJ,1999SchCzApJ}), and boxcar basis functions (e.g. Box Least Squares \citep{Kovacs_BLS}). 
Other periodograms attempt to minimize the dispersion of datasets in phase space. 
These algorithms include phase dispersion minimization \citep{stellingwerf1978PDM}, the phase binning variant of AOV~\citep{1989SchCzMNRAS,schwarzenberg1997correct}, as well as entropy methods such as Shannon entropy \citep{Cincotta_Shannon_Entropy} or the  commonly used conditional entropy \citep{Graham_CE}.
Because many period finding algorithms search for waveforms of particular shape, they may miss weak signals with disparate structure to the search criteria. Because of this, several period detection algorithms are sometimes 
used together, see, e.g., \citet{2021coughlinMNRAS}.

In this paper, we describe an algorithm, Fast Periodicity Weighting (FPW), for detecting 
periodic sources having arbitrary phase-folded waveforms. 
This paper provides a simple introduction to the algorithm, discusses its basic features, describes its implementation, and provides links to code.
A more detailed discussion will be provided in a second paper (Paper II), which will include a description of the Gaussian Process framework for deriving algorithms such as FPW. Paper II will also evaluate the statistics of the algorithm, and provide quantitative comparisons with other frequently used periodicity detection algorithms.

FPW is a phase-binning algorithm and is therefore waveform agnostic.
While FPW is not quite as sensitive as the Lomb-Scargle algorithm for pure
sinusoids, it is capable of detecting a much wider range of waveforms, including those of eclipsing objects. 
FPW can be applied to irregularly sampled data and need not be calculated on an evenly spaced frequency grid, nor does it assume uniform noise in each flux measurement. 


In Section \ref{sec:algorithm} we define the FPW algorithm and discuss its characteristics.
In Section \ref{sec:application} we provide examples of the performance of the algorithm for various sources using light curves derived from ZTF photometry.
In Section \ref{sec:implement} we discuss implementation of the algorithm on various compute platforms including timing results.
Links to both CPU and GPU code are provided.
Section \ref{sec:conclusion} gives a
brief summary and discussion.

\section{The FPW Algorithm}
\label{sec:algorithm}

The FPW algorithm is a phase-binning algorithm in which each observation is
assigned to a phase bin depending on the time of the observation and a proposed period or frequency. The statistic calculated for each period or
frequency is:
\begin{large}
\begin{equation}
    S_{FPW} \equiv \sum_{m=1}^M \left[ 
    \frac{\left( \sum_{j=1}^{N_m}  \frac{x_{m,j}}{{\sigma^2_x}_{m,j}} \right)^2}
   {\sum_{j= 1}^{N_m} \frac{1}{{\sigma_x^2}_{m,j}} +\frac{1}{\alpha^2} } \right] ~,
   \label{eq:phasesum}
\end{equation}
\end{large}
\noindent
where $M$ is the number of phase bins, $N_m$ is the number of observations in the
$m$th phase bin, $j$ is the index running over the observations in a phase bin,
$x_{m,j}$ is the mean-subtracted magnitude of the $j$th photometric observation in the $m$th phase bin, and ${\sigma_x^2}_{m,j}$
is the squared observation error of $x_{m,j}$. 
The parameter $\alpha$ represents a prior on the expected variation of the variable source, and has little effect on periodicity detection. (If it is too small, it suppresses detection -- if it is too large, the variable component takes all the flux it can, even though some should be apportioned to the noise component).
See the discussion in Appendix \ref{sec:derivation}.
The denominator in the square brackets normalizes the squared sums of
the weighted observations. 
The quantity $S_{FPW}$ is computed for each period over the range of
periods to be searched, with the result being the distribution of $S_{FPW}$ versus period or frequency, i.e., the FPW periodogram.

Equation \ref{eq:phasesum} was derived using a Gaussian
Process (GP) framework, and is formally an expression that represents the difference in $\chi^2$
between the assumption of a phase binned periodic source at the trial frequency and that 
of a constant source at the same frequency.
We provide a derivation in Appendix \ref{sec:derivation}.

In the limit of large $\alpha$, Eq. \ref{eq:phasesum} is simply the weighted sum over phase bins
of the square of the sum of the weighted observations in each phase bin.
In that sense, Eq. \ref{eq:phasesum} is simply an optimized expression for the significance of 
periodicity in a phase-folded mean-subtracted time series.
Although it is a very simple expression, we have not found Eq. \ref{eq:phasesum} in published literature on periodicity detection.  
While expressions with some similarities have appeared in other algorithms, many current astrophysical analyses use either searches with sine/cosine bases such as Lomb-Scargle, or box-car bases, such as Box Least Squares (BLS). 
FPW covers all types of waveforms, albeit with somewhat reduced sensitivity compared to
those tailored for a specific type of waveform.
See Section \ref{sec:application} for a comparison of FPW to Lomb Scargle.

The FPW algorithm allows
the photometric errors to vary from observation
to observation and the errors can be determined using available data other
than the source being analysed for periodicity.  
In the case of
ZTF, the field of view of 47 square degrees \citep{2019BellmPASP,Masci_ZTF_Survey} is divided into
64 quadrants (each corresponding to a readout channel) and all sources in a quadrant are analysed together. 
The errors in individual flux measurements are estimated using the
data from a collection of sources in the quadrant and account for variations in the point spread function and other systematic
effects common to all sources in a given quadrant.

In contrast to FPW, in the phase binning variant of AOV \citep{1989SchCzMNRAS,schwarzenberg1997correct}, the
errors (variances) of the individual measured signals are generally not assumed to be known. To remedy this, AOV uses the ratio of two quantities, both estimators of
the unknown variance.  The ratio of the two quantities then  eliminates the
dependence on the unknown variance\footnote{In \cite{schwarzenberg1997correct}, the quantity $s_1^2$ is the sum of the square of
the mean signal in each phase bin and the quantity
$s_2^2$ is the sum of the squared deviation from the mean in each bin. In the phase binning variant of AOV, the ratio
$s_1^2/s_2^2$ is used to construct a quantity independent of the sample
variance $\sigma^2$.}.
However, this increases the computational
complexity of the algorithm and does not make use of the additional information
on the variance of the source that may be available from the observation of
other nearby sources. 
In particular, FPW can weight individual observations according to the uncertainty estimate of each.\footnote{A coded version of AOV exists (AOVW in 
\url{https://users.camk.edu.pl/alex/\#software}) 
that includes observation errors.  However, we could not find an expression for this variant of AOV in the literature and a formal comparison with FPW has not been made.}

We note that the Bayesian Block formalism of Scargle et al.\,\citep{Scargle_Periodogram} can be used to derive a formula for
periodicity searches that is equivalent to Eq.\,\ref{eq:phasesum}.
The Bayesian Block formalism discusses binning of data in a very general way and discusses numerous applications, for example, the analysis of bursts, flares, and triggers. 
While it does not explicitly discuss binning into phase bins for periodicity detection, the formalism can be applied to phase
binning, and the block log likelihood given in their Eq. 41 is
equivalent to the contribution to \SFPW\ from an individual phase bin.
Because the log likelihood for an ensemble of blocks is equal to the sum of the individual block log likelihoods, the expression
for the Bayesian Block log likelihood for phase binning is equivalent
to Eq.\,\ref{eq:phasesum} of the FPW algorithm.
An interesting feature of the Bayesian Block formalism is that it determines the optimal binning strategy using unequal blocks.  Our current implementation of FPW assumes equal bin widths, but this is not intrinsic
to the algorithm and a Bayesian Block refinement of bin widths could be
considered, e.g., as a post-processing step for selected peaks in the
frequency distribution as determined by FPW.

While FPW is different from the commonly used Phase Dispersion Minimization (PDM) 
algorithm of Stellingwerf \citep{stellingwerf1978PDM}, we note that that paper
briefly mentions an alternative method from the 1924 book by Whittaker and Robinson \citep{whittaker1924calculus}.
That method searches over period for the maximum variances of the bin means. This appears
to be similar in spirit to the FPW algorithm that we describe in this paper.
FPW also has similarities to the Epoch Folding (EP) algorithm of Leahy et al. \citep{Leahy1983ApJ}, in particular equation B1. 
However, the weighting of measurements is treated differently.

An extensive comparison of FPW with other periodicity detection algorithms will be provided in Paper II.

\section{Use of FPW}  
\label{sec:application}

A key parameter of FPW is the number of phase bins, $M$.
Choice of $M$ determines the relative sensitivity of the algorithm to various
types of waveforms, e.g., sinusoidal, eclipsing, or complex.
For small $M$, e.g., $M=4$ or $5$, the FPW algorithm is most sensitive to sinusoidal
signals.  For larger $M$, FPW is sensitive to increasingly more complex waveforms, as
well as to eclipsing sources with eclipse duty cycles of $\sim1/M$ or greater.
As discussed in Section \ref{sec:multiple}, $S_{FPW}$ for multiple values of $M$ can be computed with little additional computation compared to a single value of $M$.

\begin{figure}[htbp]
    \centering
    \includegraphics[width = 0.7\textwidth]{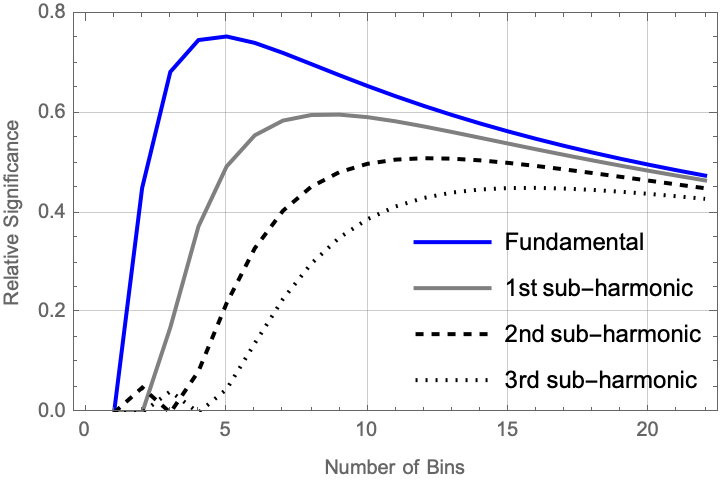}
    \caption{{\bf Significance of FPW relative to Lomb-Scargle for a sinusoidal waveform.}
    This is the predicted response of the FPW algorithm to a binned sinusoidal signal sampled at the correct period, averaged over the
    phase offset of the sinusoid. 
    The response is shown relative to the significance of detection from a $\chi^2$ fit of sinusoidal basis functions to a pure sinusoid, e.g., by an algorithm such as the Lomb-Scargle algorithm, and  
    is multiplied by a factor which accounts for the increasing number of degrees of freedom (DOF) as the number of bins is increased (DOF = $M-1$, where $M$ is the number of bins). 
    This factor accounts for the difference in the cumulative tail probability for $\chi^2$ for different DOF, normalized to a cumulative tail probability of $2 \times 10^{-9}$ for
    $\chi^2$ with 2 DOF.  
    The expression for the relative significance is discussed in Appendix \ref{sec:comparison}.
    The significance of the fundamental and sub-harmonics depend on $M$ explicitly in
    only a limited number
    of cases, principally
    $M=2$, for the fundamental, as well as a few specific values of $M$ for
    sub-harmonics, namely  $M = 2 \times (h+1)$, where $h$ is the order of the sub-harmonic. 
    }
    \label{fig:bin_dependence}
\end{figure}

\subsection{Response of FPW to a Sinusoidal Waveform}
The response of FPW to a pure sinusoid is shown in
Figure \ref{fig:bin_dependence}.
The curves are expressed relative to the response of an algorithm, such as Lomb-Scargle, that calculates
the difference between a chi-square
fit to a sinusoid and a fit to
a constant source.
Lomb-Scargle is distributed as $\chi^2$ with
two degrees of freedom\,\citep{1982ScargleApJ,VanderPlas_UnderstandingLS}.
Two competing factors combine to make FPW somewhat less sensitive to
pure sinusoids than Lomb Scargle.
Larger values of $M$ approximate a sinusoid better and thus have higher
$S_{FPW}$, but a higher $S_{FPW}$  is required to achieve the same random false alarm
rejection level as $M$ increases, i.e., the number of
degrees of freedom increases.
This is shown in Figure \ref{fig:bin_dependence}, which
shows a maximum in the fundamental peak for FPW for 4 or 5 phase bins.
Figure \ref{fig:bin_dependence} has been
calculated for a random false alarm rejection
level of $2 \times 10^{-9}$, a level useful
for searches for periodic sources in
ZTF, involving a grid of O($10^6$) frequencies and O($10^9$) sources.
The relative significance is not a strong function of the false alarm 
rejection level.  
We use it here as a simple approximate measure of relative efficiency of periodicity
detection.  See \citet{2008BaluevMNRAS,VanderPlas_UnderstandingLS} for more detailed
discussion.

Figure \ref{fig:bin_dependence} indicates that
the maximum significance ($S_{FPW}$ value) to a pure sinusoid is obtained
for $M=4$ or $5$ and that the significance is
about 75\% of what would be
obtained by a Lomb-Scargle algorithm
optimized for pure sinusoids, taking into account the larger number of degrees
of freedom (DOF) for FPW.
Observationally, the maximum detection distance to an astrophysical source with sinusoidal periodic waveform 
is therefore reduced by about 7.5\% for FPW compared to Lomb-Scargle.

\subsection{Sub-harmonics in the FPW Periodogram}
\begin{figure}[ht!]
    \centering
    \includegraphics[width = 1.0 \textwidth]{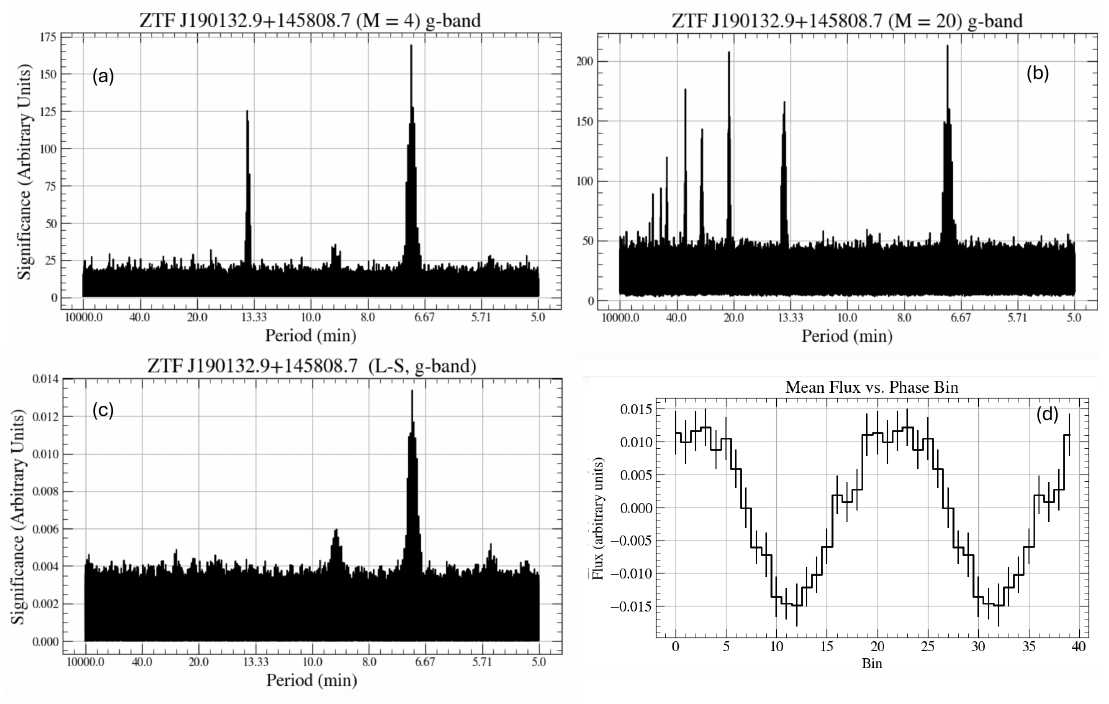}
    \caption{{\bf FPW analysis of g-band photometric data for ZTF J190132.9+145808.7 \citep{caiazzo2021smallWD}.} This source has a
    rotation period of 6.94 min and a quasi-sinusoidal phase-folded waveform.
    (a): FPW analysis with $M=4$ showing the 2 prominent peaks.  (b): FPW analysis with $M=20$, showing peaks at numerous sub-harmonics. (c): Lomb-Scargle analysis of same source. 
    (d): Two cycles of the phase-folded light curve evaluated at the frequency peak corresponding to 6.94 minutes.}
    \label{fig:numbin_comparison}
\end{figure}

FPW typically has significant peaks in the
$S_{FPW}$ periodogram at sub-harmonics of the true
period.  This is quantified in Fig. \ref{fig:bin_dependence} and illustrated in 
Figure \ref{fig:numbin_comparison}.
Figure \ref{fig:bin_dependence} plots the
magnitude of the periodicity significance at the fundamental and sub-harmonics versus the number of bins, $M$, for the case of a pure
sinusoidal waveform. 
The ratio of the height of
the peak at the true frequency to the height of peaks at sub-harmonics depends
on the shape of the binned phase-folded waveform.  
Figure \ref{fig:numbin_comparison} illustrates this feature of multiple harmonics.
The figure shows analysis of 5 years of ZTF data for the source ZTFJ190132.9+145808.7,
a rapidly rotating, highly magnetic white dwarf (WD) with a 
period of 6.94 minutes, discussed in \citet{caiazzo2021smallWD}.
The phase binned folded light curve of the source is quasi-sinusoidal. The upper left hand side of Fig. \ref{fig:numbin_comparison} shows the results
of the \FPW\ analysis of the source for the case of $M=4$.  Two significant peaks are visible,
one at the fundamental rotation period of 6.94 min, and one at the 1st sub-harmonic.
The upper right hand side of Fig. \ref{fig:numbin_comparison} is the result for the case of $M=20$, showing the numerous additional peaks located at sub-harmonics of the primary frequency.  
The lower left hand side of Fig. \ref{fig:numbin_comparison} shows the result of the comparable Lomb-Scargle analysis having only one significant peak.

\begin{figure}[ht!]
    \centering
    \includegraphics[width = 1.0 \textwidth]{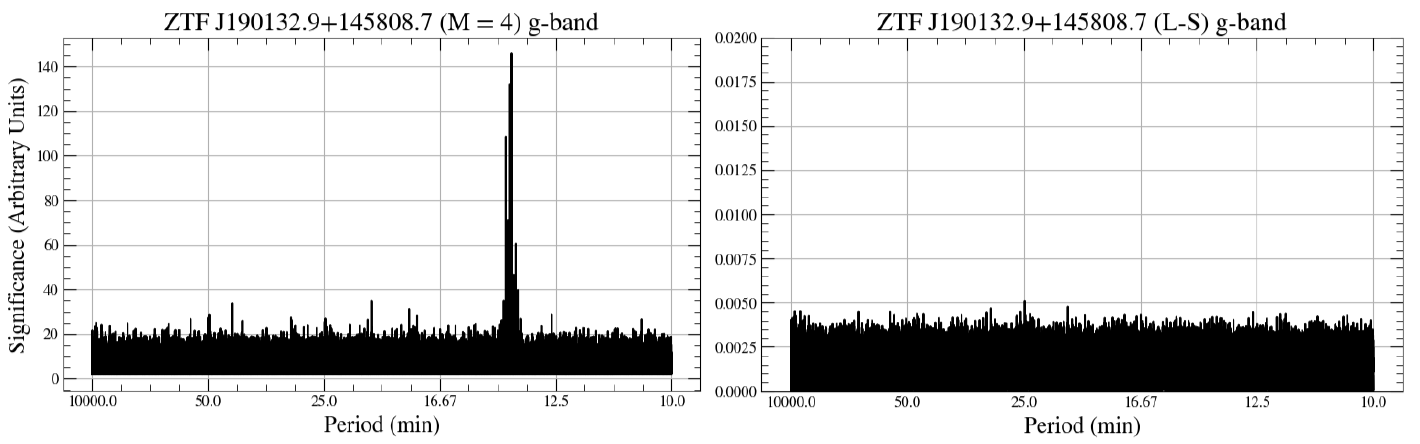}
    \caption{{\bf FPW analysis of ZTF J190132.9+145808.7 \cite{caiazzo2021smallWD} as in
    Fig. \ref{fig:numbin_comparison}
    but with a short period cutoff
    of 10 minutes.}
    Left: FPW analysis with $M=4$ showing a peak at the first sub-harmonic. Right: comparable analysis with Lomb Scargle.  No peak is detected.  The sub-harmonic sensitivity of FPW extends a factor of $M/2$ beyond the nominal period cutoff for even $M$.}
    \label{fig:sub-period}
\end{figure}

Like some other algorithms, e.g., conditional entropy \citep{Graham_CE},
FPW has the advantage of being sensitive to periodicity at periods shorter than the nominal shortest period
being evaluated in the periodogram. 
Figure \ref{fig:sub-period} illustrates this.  On the left is the FPW detection of the 6.94 minute rotating WD calculated using a short period limit of 10 minutes, i.e., longer than the true period of 6.94 minutes.
The periodic source is clearly detected due to the presence of a peak at the first sub-harmonic, albeit at a lower significance than it would be if detected at the
fundamental period.
On the right is the same analysis using Lomb-Scargle with a 10 minute short period limit for which no significant peak is detected. While the sensitivity in detection of the sub-harmonic peaks in FPW is 
not as good as for the primary peak, Fig.\,\ref{fig:bin_dependence} indicates
that sensitivities may be comparable, particularly for larger values of $M$.
Sensitivity at periods shorter than the formal period search limit can lead to a significant speed up of the algorithm as the periodicity computation is
often dominated by the number of frequency grid points at the shortest periods.

\subsection{Response of FPW to Eclipsing Sources}

\begin{figure}[ht!]
    \centering
    \includegraphics[width = 0.7\textwidth]{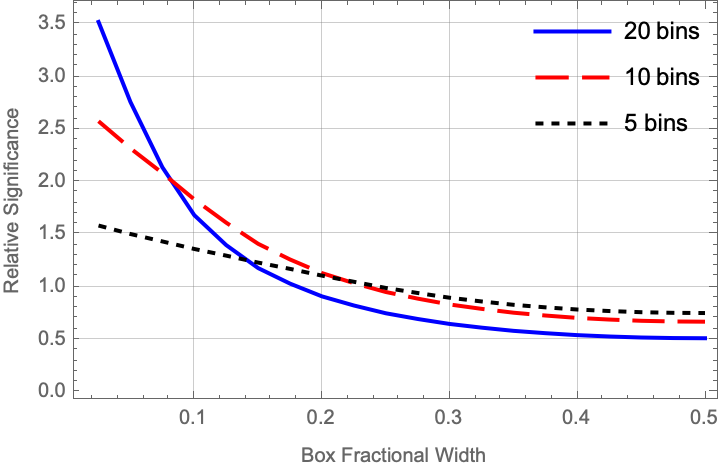}
    \caption{{\bf Significance of FPW relative to Lomb-Scargle for a box-like waveform.}
    The relative significance is \SFPW\ divided by the $\chi^2$ as
    obtained from a Lomb-Scargle style algorithm, modified by
    the difference in the $\chi^2$ needed to obtain a false alarm rejection level of $2 \times 10^{-9}$ for both Lomb-Scargle and
    FPW with the indicated number of bins (See Appendix \ref{sec:comparison}).
    For FPW with $M=20,10,5$ (which can be efficiently computed together), at least one is better than Lomb-Scargle for box
    widths up to about 25\% of the period.
    The relative significance depends on the phase offset of the leading edge of the box waveform in the phase-folded light curve, particularly for box widths comparable to or wider than a bin width. The curves are averaged over phase offset.
    }
    \label{fig:box_comparison}
\end{figure}

One of the primary reasons for the choice of FPW for ZTF periodicity analysis is its improved response to eclipses and complex waveforms relative to the Lomb-Scargle algorithm,
while still retaining acceptable sensitivity to sinusoidal sources. 
To quantify the improvement of FPW over Lomb-Scargle for eclipses, we
approximate an eclipse by a box-like function. 
Figure \ref{fig:box_comparison} shows the response of FPW to box-like functions of different widths relative to the response of a Lomb-Scargle algorithm, taking into account the difference in DOF.
The calculations have been averaged over phase offset of the box waveform.
The figure shows that for a narrow box width of 1/20th of the total period, an FPW analysis using 20 bins produces a relative significance that is approximately 3 times better than Lomb-Scargle after correcting for the relative number of degrees of freedom ($19$ for FPW and $2$ for Lomb-Scargle). 
The detection distance for FPW is consequently 30\% larger than for Lomb-Scargle for sources with eclipse duration $\le 1/20$ of the waveform.

A commonly used algorithm for eclipse detection is the box least squares (BLS) algorithm
\citep{Kovacs_BLS}.
If BLS is used in a phase-binning mode with the same number of bins
as FPW, then the equivalent BLS statistic for a narrow eclipse is
$\chi^2$ distributed with one DOF (the bin in which the eclipse occurs) while the FPW statistic \SFPW\ is distributed as $\chi^2$
with $M-1$ DOF.  
For $M=20$ and a false alarm rejection level of $2 \times 10^{-9}$ this implies that BLS has about twice the 
significance of FPW for a source with narrow eclipses.
We note however, that because FPW computes the mean signal level
in each bin, it would be straightforward to compute the full BLS statistic
in addition to the FPW statistic as part of the phase bin analysis,
albeit with larger computational complexity.
\begin{figure}[ht!]
    \centering
    \includegraphics[width = 1.0 \textwidth]{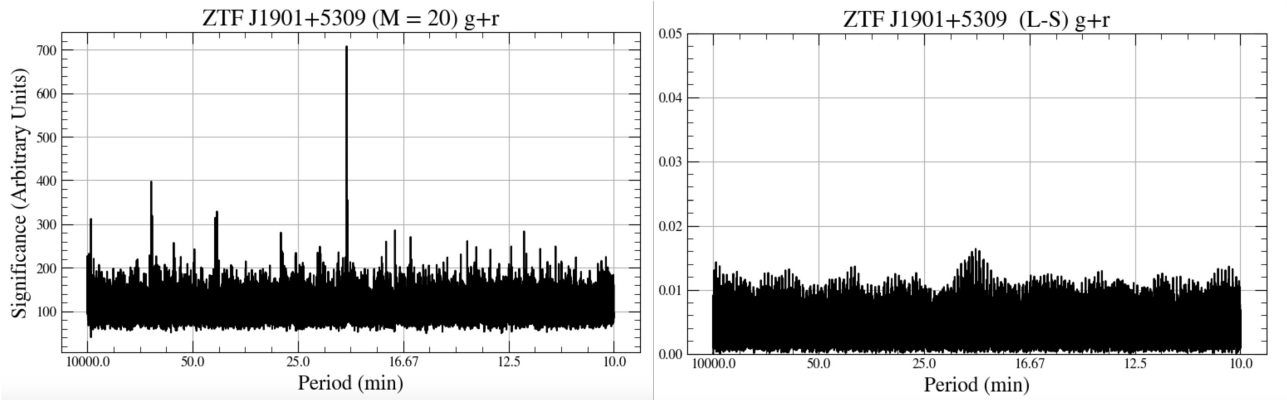}
    \caption{{\bf Analysis of the eclipsing source ZTF~J1901+5309} \citep{burdge2020systematic}. Left:  FPW analysis. 
    The source exhibits narrow primary and secondary eclipses, leading to a strong detection at half the period, i.e., at 20.3 minutes. Right: Lomb-Scargle analysis of the same data set. A broad weak peak is seen at 20.3 minutes. }
    \label{fig:eclipse}
\end{figure}

Figure \ref{fig:eclipse} compares the results of FPW and Lomb Scargle for the eclipsing double white dwarf binary, ZTF~J1901+5309 \citep{burdge2020systematic}.  The source
has an orbital period of 40.6 minutes and prominent primary and secondary eclipses, leading to a strong detection at 20.3 minutes, one half of the orbital period.  
As seen in Figure \ref{fig:eclipse}, FPW shows a sharp prominent peak at 20.3 minutes.  
Lomb-Scargle also exhibits a peak at 20.3 minutes, but broad and much less prominent, unlikely to be detected in a blind periodicity search.
The
figure was produced using a truncated data set of
50 days, so that the periodogram peaks were closer to the detection threshold to
ilustrate the advantage of FPW over Lomb Scargle for eclipsing sources.
In the full 5-year ZTF data set both FPW and Lomb-Scargle detect a strong peak at 20.3 minutes.

\subsection{Calculation of FPW for Multiple Numbers of Bins}
\label{sec:multiple}
Little additional computation is required to calculate Eq.\,\ref{eq:phasesum} for values of $M$ that are integer divisors of a maximum number of bins, $M_{\rm max}$.  They simply involve combining sums already computed for the numerator and denominator of Eq.~\ref{eq:phasesum}. 
For example, choosing $M_{\rm max}=20$ allows straightforward calculation of $M=20,10,5,4,$ and 2.  The pre-computed sums may be combined with different offsets (there are $k$ choices of offset for $M=M_{\rm max}/k$) and all of these are easily computed. 
In this case, $M=20$ will provide good sensitivity for light curves with features
with duty cycle $<1/20$, such as eclipsing sources, while $M=4$ or $5$ will provide good sensitivity for sinusoids both at the fundamental and the 1st sub-harmonic. The case of $M=10$ will be useful for intermediate 
complexity waveforms.
The significance of peaks in the spectrum of $S_{FPW}$ will decrease with increasing $M$ because of the extra degrees of freedom that waveforms can take on.  
If the waveforms are relatively smooth on the scale of $1/M$ of the folded light curve,
then further increasing the number of bins will result in decreasing significance for detection of periodicity.  Because computation time is only weakly dependent on $M_{\rm max}$, it may be sensible to run with a high $M_{\rm max}$ (say, 40 or 60) making the $M$ of interest (20 or 10 or 4) available with multiple bin offsets.

\subsection{Pre-processing of Data and Post-processing of Results}
While not part of the FPW algorithm, pre-processing of the data and post-processing of the calculated 
periodograms are important parts of the overall ZTF analysis.
Here, we briefly describe some of the considerations for pre- and post-processing,
leaving details for a future paper discussing results of the full processing of 5 years of ZTF data with the FPW algorithm.

Pre-processing of data includes removal of long-period baseline variations followed by subtracting the 
average value of the observations to produce a zero-mean sequence of data points.
The baseline correction step is important to reduce false positives in the FPW periodicity analysis.
As explained further in Appendix \ref{sec:phase_entropy}, monotonic trends in the baseline combined with
regularities in the sampling times can produce spurious peaks in the periodogram at the characteristic 
period of the regularities. Producing a zero-mean data set as input to the FPW algorithm simplifies the
analysis and removes one degree of freedom from the resulting periodogram. 

Post-processing of the results of the FPW analysis typically involved selection of some number of the
highest amplitude peaks in the periodogram. 
As discussed in Appendix \ref{sec:phase_entropy}, the periods of these peaks can then be compared to 
lists of possible spurious peaks, e.g., by using phase entropy.

\section{Implementation of FPW}
\label{sec:implement}
We provide a Python package written in \texttt{C++} and ported to python with \texttt{Cython} for computing FPW on individual light curves as well as batches of light curves. The complete program is available freely under the MIT license. The python wrapper takes in a float64 array of timestamps, a float64 array of flux values, a float64 array of flux errors, a float64 array of test frequencies, and an integer number of bins. A basic example and testing Jupyter Notebook is provided with the package source code.
In the rest of the section we outline the base code that the package computes (\ref{lst:FPW_Functions}).
The \texttt{C++} source code is provided in Appendix \ref{sec:code}.
The code requires only packages included in the \texttt{C++} standard libraries. The source code and examples are available freely at \url{https://github.com/Sewhitebook/FPW}. Highly parallelized GPU code is currently under development: the source code for which is available upon request to the authors.


We list the \texttt{C++} source code for FPW in listing \ref{lst:FPW_Functions}. The code loops over every test frequency in the provided array. For each frequency, \texttt{f}: 
\begin{itemize}
    \item The \texttt{makeIndices} function (lines 1 \--- 11) computes the bin that each timestamp in \texttt{times} falls into. We achieve this efficiently by subtracting the integer part of the phased timestamp, and multiplying that fraction by the number of bins. The integer part of this number is then the zero-indexed phase bin,
    which is stored in the \texttt{ind} array.
    \item The \texttt{deltaChi2} function (lines 13 \--- 40) computes the \SFPW \, statistic for the test frequency corresponding to the \texttt{ind} array.
    The denominator of Eq.~\ref{eq:phasesum} is contained in \texttt{VtCinvV}, with the numerator tracked by \texttt{ytCinvV}.
    The array \texttt{ind} has length $N_\mathrm{data}$, listing which bin each data point falls into.
    \item Lines 23 \--- 28 use the indices in the array \texttt{ind} to accumulate the error and data weighted by error into the respective bins.
    \item Lines 30 \--- 33 then handle the outer loop of Eq. \ref{eq:phasesum}, i.e, the loop over bin indices.
\end{itemize}
     Finally, \texttt{runFPW} (lines 42 \--- 61) is a simple wrapper function to compute the above functions over every test frequency and store the results. We also provide a batch function for running FPW on multiple light curves that share the same timestamps. Batching is advantageous because the \texttt{makeIndices} function only needs to be computed once for the batch.

     \begin{figure}
    \centering
    \includegraphics[width=0.5\linewidth]{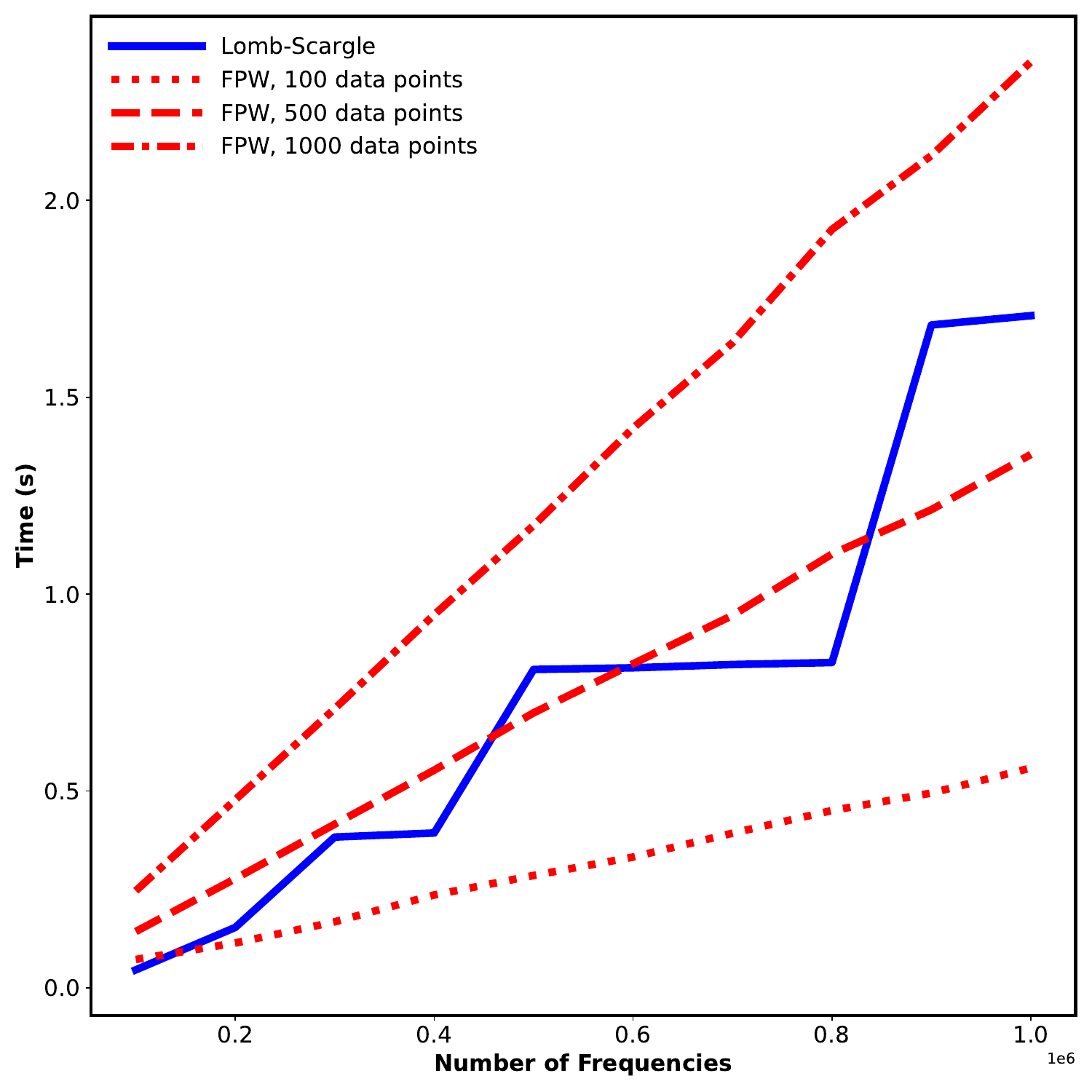}
    \caption{Benchmarking results from FPW and Astropy Lomb-Scargle in `fast' mode on a 3.10 GHz Intel i9-9960X processor. Red lines show FPW for different numbers of data points. The blue curve shows Lomb-Scargle with 1000 data points. As shown in \cite{Press_Rybicki_LS}, the fastest implementation of Lomb-Scargle utilizing FFTs does not depend strongly on the number of data points. We note that the constant factor on the Lomb-Scargle implementation allows FPW to be faster up to about 500 data points.}
    \label{fig:fpw_benchmark}
\end{figure}

We benchmark the FPW algorithm against the fastest astropy implementation of Lomb-Scargle with no normalization \citep{astropy:2013, astropy:2018, astropy:2022}. To do this, we generate Gaussian noise signals and run each algorithm over the dataset on a Linux machine with a 3.10 GHz Intel i9-9960X processor. The computation time comparison of FPW and Lomb-Scargle is in Figure \ref{fig:fpw_benchmark}. The computational cost of increasing the number of phase bins is concentrated in the outer sum of equation \ref{eq:phasesum}. Since the number of data points should be at least an order of magnitude higher than the number of phase bins to maintain statistical rigor, the cost associated with changing the number of bins is additive upon, and negligible compared to, the cost scale associated with the number of test frequencies and the number of data points. As seen from equation \ref{eq:phasesum}, FPW is $\mathcal{O}(N_\mathrm{freqs} N_\mathrm{data})$ complex. We compare this to the fastest astropy implementation of Lomb-Scargle, which exploits FFTs \citep{Press_Rybicki_LS} to achieve $\mathcal{O}(N_\mathrm{freqs} \, \log{N_\mathrm{freqs}})$ complexity. We note, however, that this implementation of Lomb-Scargle has a large constant time, which makes FPW faster within a common regime of analysis. 
For datasets with $\lesssim 500$ points, up to $\sim 1e7$ frequencies FPW is faster than Lomb-Scargle per source.

\section{Summary and Conclusions}
\label{sec:conclusion}
We describe a phase binning algorithm (FPW) for detecting periodicity in non-uniformly sampled time series data.
We demonstrate that the algorithm has relatively good sensitivity to periodic sinusoidal waveforms when compared to the widely-used Lomb-Scargle algorithm, but has considerable better sensitivity to periodic eclipses compared to the Lomb-Scargle algorithm.  
We have therefore chosen it for the periodicity analysis of ZTF data, which can have complex
waveforms ranging from sinusoidal wave forms, to box-like wave forms, to narrow eclipses, and to more complex combinations of several waveform components.  
We provide examples of applications of the algorithm to ZTF data from a short period rotating white dwarf and a compact white dwarf binary.

We provide code for the implementation of the FPW algorithm and show that the computational efficiency of the 
algorithm is comparable to the widely-used Lomb-Scargle algorithm. 

\begin{acknowledgments}
{\bf Acknowledgments: } 
This work supported in part by a NASA LISA Preparatory Science Grant, 80NSSC21K1723.
This work uses observations obtained with the Samuel Oschin 48 inch 
Telescope at the Palomar
Observatory as part of the Zwicky Transient Facility project.
Major funding for ZTF has been provided by the U.S. National Science
Foundation under grant No.~AST-1440341 and by the ZTF
partner institutions.  DF was partially supported by the National Science Foundation under Cooperative Agreement PHY2019786 (The NSF AI Institute for Artificial Intelligence and Fundamental Interactions).
The observational data was provided by the NASA/IPAC Infrared Science Archive, which is funded by the National Aeronautics and Space Administration and operated by the California Institute of Technology.
We acknowledge useful comments from colleagues on the draft manuscript, in particular those from Ilaria Caiazzo, Matthew Graham, and Daniel Warshofsky.
\end{acknowledgments}
\clearpage


\appendix
\newcommand{\deltachisq}{$\Delta \chi^2$}

\section{Derivation of FPW}
\label{sec:derivation}

The FPW method was originally derived as a way to marginalize over baseline drifts that can cause troublesome aliasing near a period of one sidereal data.  Because this paper focuses on shorter periods, we use a faster version of FPW that neglects the baseline drift term and therefore runs an order of magnitude faster.  We begin by deriving the general case and then removing the baseline drift component to obtain Eq. \ref{eq:phasesum}.

\subsection{The model}

We model the mean-subtracted light curve samples as the sum of a baseline drift, a periodic signal, and noise. Let the mean-subtracted flux (or magnitude) at time $t_i$ be denoted by $x_i$. The model assumes:

\begin{equation}
x_i = \sum_k c_k b_{ik} + y_{m(i,f)} + \epsilon_i \, ,
\end{equation}
where $b_{ik}$ represents the $k^th$ polynomial basis function for the baseline evaluated at time $t_i$, $c_k$ are the polynomial coefficients,  $y_{m(i,f)}$ is the periodic signal corresponding to phase bin $m(i,f)$ (which depends on frequency $f$), and $\epsilon_i$ is the noise realization drawn from a Gaussian with standard deviation $\sigma_i$.  The phase bin index is
\begin{equation}
    m_i =  {\rm floor}(N_{\rm bin} [(t_i - t_{\rm ref})/\tau\mod 1]) 
\end{equation}
for some reference time, $t_{\rm ref}$, which is often taken to be the time of the first observation.  The period $\tau=1/\nu$, and $m$ is zero-indexed, ranging from 0 to $N_{\rm bin}-1$.

The covariance matrix for each component is either low rank (can be expressed as a product of a small number of basis vectors times its transpose) or diagonal (in the case of the noise). 
\begin{itemize}
    \item $C_B = B B^T$: the covariance matrix for the baseline model, where each column of $B$ represents a polynomial evaluated at the observation times.  In the absence of any specific prior information about the baseline drift, a Chebyshev polynomial basis would be appropriate.
    \item $C_P = P P^T$: the covariance matrix for the periodic signal, where each row of $P$ corresponds to an observation, and contains a single $\alpha$ in the column corresponding to the relevant phase bin, and 0 elsewhere.  In other words, our prior on the signal in each phase bin is a Gaussian with mean 0 and standard deviation $\alpha$.
    \item $C_\sigma = \text{diag}(\sigma_i^2)$: the noise covariance matrix, where the noise variance is $\sigma_i^2$ for each observation $i$.
\end{itemize}

The difference in chi-square, $\Delta \chi^2$, between the baseline+periodic signal+noise model and the baseline+noise model is given by:
\begin{equation}
\Delta \chi^2 = X^T (C_B + C_P + C_\sigma)^{-1} X - X^T (C_B + C_\sigma)^{-1} X
\end{equation}
where $X$ is the column vector containing the $x_i$ values. 

\subsection{Efficient Computation Using Sherman-Morrison}

Calculating $\Delta \chi^2$ directly for each trial frequency $f$ would be computationally expensive, especially given the large number of frequencies to sweep over. To mitigate this, we apply the Sherman-Morrison formula to efficiently update the inverse when the periodic signal model is added.

First, precompute $(C_B + C_\sigma)^{-1}$, which is independent of the trial frequency:
\begin{equation}
C_{B\sigma}^{-1} = (C_B + C_\sigma)^{-1}
\end{equation}

Next, apply the Sherman-Morrison formula to account for the addition of $C_P$:
\begin{equation}
(C_B + C_P + C_\sigma)^{-1} = C_{B\sigma}^{-1} - C_{B\sigma}^{-1} P (I + P^T C_{B\sigma}^{-1} P)^{-1} P^T C_{B\sigma}^{-1}
\end{equation}

The difference in chi-square then simplifies to:
\begin{equation}
\Delta \chi^2 = - X^T C_{B\sigma}^{-1} P (I + P^T C_{B\sigma}^{-1} P)^{-1} P^T C_{B\sigma}^{-1} X
\end{equation}

This expression allows us to efficiently compute $\Delta \chi^2$ for each trial frequency by reusing the precomputed inverse $C_{B\sigma}^{-1}$ and updating it based on the matrix $P$, which depends on frequency.

\subsection{Simplified Formulation Without the Baseline}

In this paper we are especially interested in short periods where the baseline drift is insignificant, so we further simplify the computation by modeling and subtracting the baseline, rather than marginalizing over it with the above formula.  This runs the risk of aliasing power onto the observing cadence, but speeds up the computation substantially.  For a pre-subtracted baseline, we can neglect the polynomial term ($C_B = 0$). In this case, we are left with the noise covariance $C_\sigma = \text{diag}(\sigma_i^2)$ and the periodic signal covariance $C_P$. The chi-square difference becomes:
\begin{equation}
\Delta \chi^2 = X^T (C_P + C_\sigma)^{-1} X - X^T C_\sigma^{-1} X
\end{equation}

Using the Sherman-Morrison formula again, with $C_\sigma^{-1} = \text{diag}\left( \frac{1}{\sigma_i^2} \right)$, we can write:
\begin{equation}
(C_P + C_\sigma)^{-1} = C_\sigma^{-1} - C_\sigma^{-1} P \left( I + P^T C_\sigma^{-1} P \right)^{-1} P^T C_\sigma^{-1}
\end{equation}

Substituting into the chi-square difference, we obtain:
\begin{equation}
\Delta \chi^2 = X^T C_\sigma^{-1} P \left( I + P^T C_\sigma^{-1} P \right)^{-1} P^T C_\sigma^{-1} X
\end{equation}

\subsection{Exploiting Sparsity for Efficient Computation}

Noting that $C_\sigma$ is diagonal and $P$ is sparse, we can reorganize the calculation for additional efficiency. Recall that each row of $P$ contains a single $\alpha$ corresponding to the phase bin for the observation time $t_i$ and frequency $f$. This expresses the prior that each phase bin value is mean zero with RMS $\alpha$.

\paragraph{Expression for $P^T C_\sigma^{-1} X$:}
Each element of $P^T C_\sigma^{-1} X$ corresponds to the sum of the observed values $x_j$ times their respective inverse variances in the respective phase bin scaled by $\alpha$:
\begin{equation}
(P^T C_\sigma^{-1} X)_m = \alpha \sum_{j \in \text{bin } m} \frac{x_j}{\sigma_j^2}
\end{equation}

\paragraph{Expression for $P^T C_\sigma^{-1} P$:}
Each diagonal element of $P^T C_\sigma^{-1} P$ is the sum of inverse variances for the observations $j$ in phase bin $m$, scaled by $\alpha^2$:
\begin{equation}
(P^T C_\sigma^{-1} P)_{mm} = \alpha^2 \sum_{j \in \text{bin } m} \frac{1}{\sigma_j^2}
\end{equation}

Thus, the chi-square difference can be written as a sum over the phase bins:
\begin{equation}
\Delta \chi^2 = \sum_{m=1}^{M} \frac{\left( \sum_{j \in \text{bin } m} \frac{x_j}{\sigma_j^2} \right)^2}{\frac{1}{\alpha^2} + \sum_{j \in \text{bin } m} \frac{1}{\sigma_j^2}}
\end{equation}

The factor $1/\alpha^2$ in the denominator is
a factor that determines the relative importance
of prior knowledge of the model versus uncertainty due to measurements.
For FPW we choose an unconstrained model for
the phase-folded waveform, and therefore prior knowledge of the model is vague, implying that
$1/\alpha^2$ is small and can be neglected. This final expression leverages the sparsity of $P$ and allows for fast computation of $\Delta \chi^2$ across many trial frequencies, as it avoids explicit matrix multiplications and instead relies on summations over the phase bins.

In future work we will reintroduce the baseline marginalization, which is expected to be important for longer period variables.  For periods much shorter than one day, the aliasing is not significant.

\section{Significance Comparison}
\label{sec:comparison}

This appendix discusses the definition of
relative significance used in Figures
\ref{fig:bin_dependence} and \ref{fig:box_comparison}.
The relative significance depends on two factors:
(1) the ratio of effective significance of FPW 
to that of a fit to sinusoidal basis functions, and (2) the ratio
of cumulative tail probabilities between $\chi^2$ distributions
with different number of degrees of freedom (DOF).

The $\chi^2$ fit to a model using a set of basis functions is given by
\begin{equation}
    \chi^2_{model} = \sum_{j=1}^{N_{tot}} \left( 
    \frac{x_j - \hat{x}(t_j)}{\sigma_j}\right)^2
\end{equation}
\noindent
where $N_{tot}$ is the total number of observations, $x_j$ is the data value at time $t_j$, $\sigma_j$ is the error on the
$jth$ data value, and $\hat{x}(t_j)$ is the model estimate at time $t_j$ using the basis functions.  
For simplicity in this calculation we assume constant errors, $\sigma_j = \sigma_x$ for all $j$.
The sinusoidal basis functions, such as used in Lomb-Scargle, are sine and
cosine.  
For FPW, the basis functions are taken to be a constant value, $\bar x_m$ for any $j$ in a
given phase bin, and $\bar x_m$ is to be computed as the weighted average of the $x_j$ in a given bin.

For a constant model with zero mean data, we estimate $x_j=0$ for all $j$.
For simplicity we also assume that the data are zero mean.
The difference between the $\chi^2$ for a given model with the data mean subtracted and a constant model at 0 is then
\begin{align}
    \Delta \chi^2 &= \frac{1}{\sigma_x^2} \left[ 
    \sum_{j=1}^{N_{tot}} \left( 
    x_j - \hat{x}(t_j)\right)^2 - \sum_{j=1}^{N_{tot}} \left( 
    x_j\right)^2
    \right]\\
    &= \frac{1}{\sigma_x^2}
    \sum_{j=1}^{N_{tot}} \left[
   \hat{x}^2(t_j)- 2 x_j\hat{x}(t_j)\right]
\end{align}
For unit amplitude sinusoidal data, $x_j = \mathrm{sin}(\phi_j)$, where $\phi_j$ is the phase of $t_j$ when phase folded on a given period with some phase offset.
For small $\sigma_x$, sinusoidal basis functions will fit the data well such
that $\hat x (t_j) \sim \mathrm{sin}(\phi_j)$ and $\Delta \chi^2$ becomes
\begin{equation}
    \Delta \chi^2_{sin} \sim - \frac{1}{\sigma_x^2} \sum_{j=1}^{N_{tot}} \mathrm{sin}^2{\phi_j}
\end{equation}

For basis functions that are a constant over a given phase bin,
as in FPW, the expression for $\Delta \chi^2$ becomes
\begin{align}
    \Delta \chi^2_{FPW} &= \frac{1}{\sigma_x^2}\left[
    \sum_{m=1}^M \left(
    N_m \ \bar x_m^2 - 2\, \bar x_m \sum_{l=1}^{N_m} x_{m,l}\right) \right]\\
    &\sim -\frac{1}{\sigma_x^2}\left[
    \sum_{m=1}^M N_m \ \bar x_m^2 \right]\\
    &\sim -\frac{1}{\sigma_x^2} \frac{N}{M} \sum_{m=1}^M \bar x_m^2 
\end{align}
\noindent
where the second line assumes small errors and large numbers of
observations so that the bin average accurately estimates the average
of $\mathrm{sin(\phi)}$ over the bin and the third line assumes a large number of observations at random times over $M$ equal width bins.

The ratio of $\Delta \chi^2$ is then
\begin{large}
\begin{equation}
\frac{\Delta \chi^2_{FPW}}{\Delta \chi^2_{sin}} \sim 
\frac{\frac{N}{M} \sum_{m=1}^M \bar x_m^2}{\sum_{j=1}^{N_{tot}} \mathrm{sin}^2{\phi_j}}
\end{equation}
\end{large}
\noindent
This quantity is calculated numerically as part of the relative
significance calculation.

The other part of the relative significance is the ratio of cumulative tail distribution functions.  The cumulative tail distribution function ($\overline{CDF} = 1 -CDF$, the complementary cumulative distribution function)
for a $\chi^2$ distribution with $k$ degrees of freedom is given by
\begin{equation}
    \overline{CDF}(X,k) = \frac{\Gamma(k/2,X/2)}{\Gamma(k/2)}
\end{equation}
where $\Gamma(k/2)$ denotes the {\it Gamma function}$, \Gamma(k/2,X/2)$ is the {\it upper incomplete gamma function}, and $X$ is the value of $\chi^2$ for which the $\overline{CDF}$ is calculated.

Assuming $k=2$ for a fit to sinusoidal basis functions, and $k=M-1$ for
a fit to a constant value in each bin (assuming zero mean), then the
relative significance of FPW to Lomb-Scargle is estimated as
\begin{large}
\begin{align}
    \mathrm{Relative\ Significance} &= \frac{\frac{N}{M} \sum_{m=1}^M \bar x_m^2}{\sum_{j=1}^{N_{tot}} \mathrm{sin}^2{\phi_j}} \times
    \frac{\Gamma((M-1)/2,X/2)}{\Gamma(1,X/2)} \times 
    \frac{\Gamma(1)}{\Gamma(\frac{M-1}{2})}\\
    &= \frac{\frac{N}{M} \sum_{m=1}^M \bar x_m^2}{\sum_{j=1}^{N_{tot}} \mathrm{sin}^2{\phi_j}} \times
    \frac{\Gamma((M-1)/2,X/2)}{\Gamma(\frac{M-1}{2})} \times 
    e^{X/2}
\end{align}
\end{large}
The relative significance is numerically calculated averaging over numerous trials of the phase offset of the $\mathrm{sin}()$ function and its bin average,
and using a value of $X$ corresponding to a cumulative tail distribution of $2 \times 10^{-9}$.
This yields the quantity shown in Figure \ref{fig:bin_dependence}.
A similar numerical calculation made using box waveforms rather than sinusoids yields the curves in Figure \ref{fig:box_comparison}.

\section{Baseline Correction and Phase Entropy}
\label{sec:phase_entropy}

Non-uniform sampling can introduce spurious peaks in the periodogram, particularly
if there are regularities in the sampling schedule, e.g., at the sidereal day period if
observations can only be done at night. 
One way of treating this is to compute the window function
of the sampling schedule, as discussed in \cite{VanderPlas_UnderstandingLS} and use the window function
to identify possible failure modes in the period determination.

Uncorrected trends in the source magnitudes can be a major source of spurious peaks in a periodogram,
e.g., on the sidereal day period or its harmonics.
We have found that subtracting long-period baseline variations significantly reduces spurious peaks
in ZTF periodicity data. However, such baseline correction can remove true variations from the source data, or the
baseline correction can be incomplete, leaving spurious peaks in the periodogram.
Upon examination, we have found that in many cases, spurious peaks in the periodogram are associated
with a non-uniform distribution of time samples in phase across the phase folded data.

A non-uniform phase distribution can be quantified by using phase entropy.
\begin{equation}
    H_{phase} = - \sum_{m=1}^M \frac{N_m}{N_{tot}}\,\mathrm{log}\left( \frac{N_m}{N_{tot}} \right)
\end{equation}
\noindent
where $M$ is the number of bins, $N_{tot}$ is the total number of observations and $N_m$ is the number of events in phase bin $m$.

For random observation times and equal bin widths, we expect approximately equal numbers of observations in each bin.
For exactly equal number of sources in each bin, $N_m/N_{tot}=-1/M$ and $H_{phase}$ = $\mathrm{log}(M)$, 
the maximum possible phase entropy given $M$ bins.
The thesis by Hutcheson \citep{hutcheson1969thesis}, provides approximations to the expectation, $E(H_{phase})$,
and the variance, $Var(H_{phase})$, for the case of equal probability of observations per bin,
\begin{align}
    E(H_{phase}) &= \mathrm{log}(M) - \frac{M-1}{2 N_{tot}} - \frac{(M-1)(M+1)}{12 N_{tot}^2} + 
    O\left( \frac{1}{N_{tot}^3} \right)\\
    Var(H_{phase}) &= \frac{M-1}{2 N_{tot}^2} + \frac{M^2-1}{6 N_{tot}^3} + 
   O\left( \frac{1}{N_{tot}^4} \right)
\end{align}

The expectation and variance of the phase entropy only depend on the sample times and not on the magnitudes
of the observations, so they need be computed only once for all sources in an image field.
This could be done for all periods as a pre-processing step before the FPW calculation for each
source in a field. 
Periods with low entropy, i.e., large negative deviation from the phase entropy expectation, can be flagged and
zeroed in the computed periodograms.
Alternatively, the expectation and variance of the phase entropy can be computed along with any
identified significant peak in the FPW periodogram.  
Peaks in the periodogram with anomalously low $H_{phase}$ can then be examined individually.
Our processing for ZTF takes the latter approach.

\begin{figure}[ht!]
    \centering
    \includegraphics[width = 1.0\textwidth]{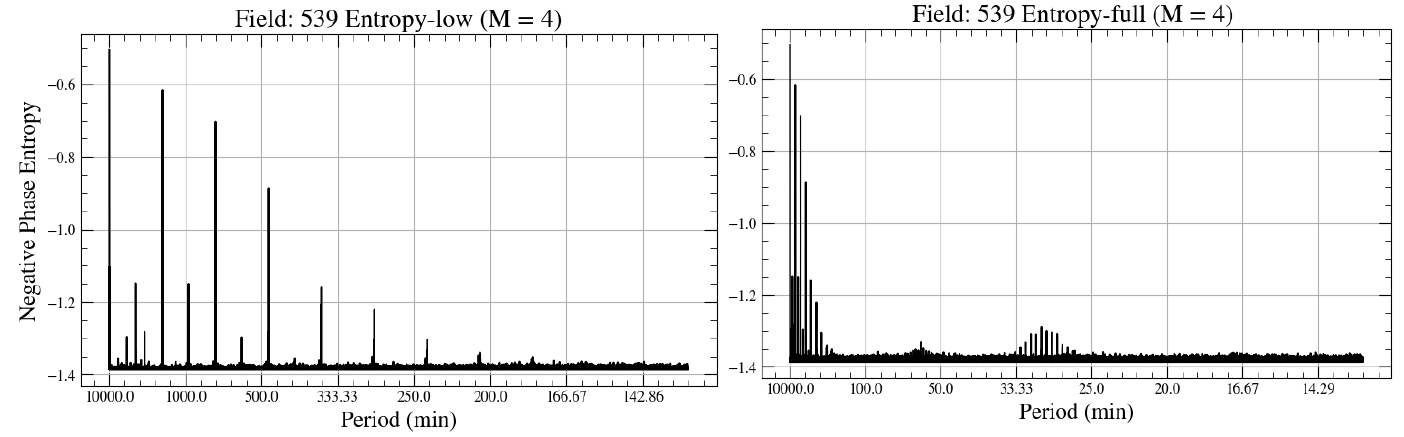}
    \caption{{\bf Phase Entropy distribution for the observation times of the ZTF field  539.}
    This is the ZTF field containing the source ZTF J190132.9+145808 discussed earlier.
    Left-hand plot: the dominant sequence of peaks are at harmonics of the approximate sidereal day frequency. Other peaks are due to
    the sub-harmonics of FPW.
    Right-hand plot: same as left-hand but also including the short period regime.
    Peaks are present at 30 and 60 minutes due to the scheduling of ZTF observations, which are blocked into 30 and 60 minute blocks
    scheduled on the hour and half hour referenced to UT.
    }
    \label{fig:phase_entropy}
\end{figure}

Figure \ref{fig:phase_entropy} shows an example distribution of phase entropy values calculated over the same grid of frequencies used to
generate the FPW periodogram.  
Periods with low entropy (high negative entropy) exhibit non-uniform distribution of time samples over phase bins.
These are candidates for spurious peaks in the computed FPW periodogram, or, e.g., the Lomb-Scargle periodogram.
As mentioned earlier, careful baseline subtraction of trends can eliminate many spurious peaks due to sampling regularities.

\newpage
\section{FPW Code}
\label{sec:code}
\begin{footnotesize}
\begin{lstlisting}[language=C, basicstyle=\footnotesize, floatplacement=tbp, caption=Listing of the basic functions to run FPW. \label{lst:FPW_Functions}]
/**
 * @brief Given N phase bins, makeIndices maps the bin index that each timestamp falls in.
 * 
 * This function calculates the phase for each timestamp and maps it to a bin index.
 * 
 * @param times Array of timestamps.
 * @param f Single frequency to be tested.
 * @param N_bins Number of phase bins.
 * @param N_dat Number of data points.
 * @return Pointer to an array of bin indices corresponding to each timestamp.
 */
int* makeIndices(double* times, const double f, const int N_bins, const int N_dat){
    int* indices = new int[N_dat];

    for(int i = 0; i < N_dat; i++){
        double phase = times[i] * f;
        phase -= static_cast<int>(phase); // Gets the fractional part of phase
        indices[i] = static_cast<int>(phase * N_bins); // The index for any given phase is the integer part of phase * N_bins
    }
    return indices; // Indices has the length of the timeseries
}

/**
 * @brief Given the indices of the timestamps, deltaChi2 computes the FPW statistic.
 * 
 * This function calculates the FPW statistic by accumulating the inverse variance
 * and the product of the inverse variance and the data for each bin.
 * 
 * @param y Array of observed values.
 * @param ivar Array of inverse variances.
 * @param ind Array of bin indices for each timestamp.
 * @param N_bins Number of phase bins.
 * @param N_dat Number of data points.
 * @return Computed FPW statistic.
 */
double deltaChi2(double* y, double* ivar, int* ind, int N_bins, int N_dat){
    double* VtCinvV = new double[N_bins]();
    double* ytCinvV = new double[N_bins]();
    double* ivar_y = new double[N_dat];

    for (int i = 0; i < N_dat; ++i){
        ivar_y[i] = ivar[i] * y[i];
    }

    double deltaChi = 0.0;
    for (int i = 0; i < N_dat; ++i){
        int index = ind[i];
        VtCinvV[index] += ivar[i]; // Accumulate the inverse variance for each bin
        ytCinvV[index] += ivar_y[i]; // Accumulate the product of the inverse variance and the data for each bin
    }

    for (int i = 0; i < N_bins; ++i){
        double Minv = 1.0 / (2 * VtCinvV[i]); //build the denominator of S_FPW out of the accumulated values
        deltaChi += ytCinvV[i] * Minv * ytCinvV[i]; // Compute the outer sum of S_FPW
    }

    delete[] VtCinvV;
    delete[] ytCinvV;
    delete[] ivar_y;

    return deltaChi;
}

/**
 * @brief Computes the relevant statistic for every test frequency.
 * 
 * This function calculates the FPW statistic for each test frequency
 * by computing the inverse variance and looping over all frequencies.
 * 
 * @param t Vector of time points.
 * @param y Vector of observed values.
 * @param dy Vector of uncertainties in the observed values.
 * @param freqs Vector of test frequencies.
 * @param N_bins Number of bins to use in the computation.
 * @return Vector of computed statistics for each frequency.
 */
vector<double> runFPW(vector<double>& t, vector<double>& y, vector<double>& dy, const vector<double>& freqs, int N_bins){
    int N_freqs = freqs.size();
    int N_dat = t.size();

    // Compute the inverse variance
    vector<double> ivar(N_dat);
    for (int i = 0; i < N_dat; ++i){
        ivar[i] = 1.0 / (dy[i] * dy[i]);
    }

\section{Comparison of FPW to Lomb Scargle}

    // Loop over all frequencies and compute the FPW statistic
    vector<double> deltaChiArr(N_freqs);
    for (int i = 0; i < N_freqs; ++i){
        int* indices = makeIndices(t.data(), freqs[i], N_bins, N_dat);
        deltaChiArr[i] = deltaChi2(y.data(), ivar.data(), indices, N_bins, N_dat);
        delete[] indices;
    }

    return deltaChiArr;
}

/**
 * @brief Computes the relevant statistic for multiple light curves sharing the same timestamps.
 * 
 * This function calculates the FPW statistic for each test frequency
 * for multiple light curves by computing the inverse variance
 * and looping over all frequencies.
 * 
 * @param t Vector of time points.
 * @param y Vector of vectors of observed values for multiple light curves.
 * @param dy Vector of vectors of uncertainties in the observed values for multiple light curves.
 * @param freqs Vector of test frequencies.
 * @param N_bins Number of bins to use in the computation.
 * @return Vector of vectors of computed statistics for each frequency for each light curve.
 */
vector<vector<double>> runFPWMulti(vector<double>& t, vector<vector<double>>& y, vector<vector<double>>& dy, const vector<double>& freqs, int N_bins){
    int N_freqs = freqs.size();
    int N_dat = t.size();
    int N_curves = y.size();

    vector<vector<double>> ivar(N_curves, vector<double>(N_dat));
    for (int j = 0; j < N_curves; ++j){
        for (int i = 0; i < N_dat; ++i){
            ivar[j][i] = 1.0 / (dy[j][i] * dy[j][i]);
        }
    }

    vector<vector<double>> deltaChiArr(N_curves, vector<double>(N_freqs));
    for (int i = 0; i < N_freqs; ++i){
        int* indices = makeIndices(t.data(), freqs[i], N_bins, N_dat);
        for (int j = 0; j < N_curves; ++j){
            deltaChiArr[j][i] = deltaChi2(y[j].data(), ivar[j].data(), indices, N_bins, N_dat);
        }
        delete[] indices;
    }
    return deltaChiArr;
}

\end{lstlisting}
\end{footnotesize}

\newpage
\bibliography{refs,sources}
\bibliographystyle{aasjournal}

\end{document}